\documentclass[%
 amsmath,amssymb,
 aps,
]{revtex4-2}

\usepackage{graphicx}
\usepackage{dcolumn}
\usepackage{bm}
\usepackage{subcaption}

\begin{document}

\preprint{APS/123-QED}

\title{Nuclear structure study using a hybrid approach of shell model and Gogny-type density functionals}

\author{Kota Yoshinaga$^{1}$, Noritaka Shimizu$^{2,3}$, and Takashi Nakatsukasa$^{2,3}$}

\address{%
$^{1}$ Graduate School of Science and Technology, University of Tsukuba, Tsukuba, 305-8571, Japan\\
$^{2}$ Center for Computational Sciences, University of Tsukuba, Tsukuba, 305-8577, Japan\\
$^{3}$ Institute of Pure and Applied Sciences, University of Tsukuba, Tsukuba, 305-8571, Japan
}

\begin{abstract}
Nuclear density functional theory (DFT) is able to reproduce
the saturation properties of nuclear matter, as well as
properties of finite nuclei.
Consequently, the DFT calculations are applicable to nuclei across a wide range of masses on the nuclear chart.
The Gogny-type density functional, which is equivalent to the mean-field calculations with finite-range density-dependent effective interactions, is a successful example.
In contrast, the shell model (configuration interaction) calculation is a powerful tool to describe nuclear structure, especially spectroscopic properties.
The shell model is able to take into account correlations beyond mean-field in a truncated model space.
In this work, we report an investigation on $sd$-shell nuclei and Ca isotopes using a hybrid approach of the shell model and Gogny-type DFT.

\begin{description}
\item[Keywords]
Nuclear density functional; shell-model calculations; $sd$-shell nuclei
\end{description}
\end{abstract}

\maketitle


\section{Introduction}

Density functional theory (DFT) is a successful approach to describing basic properties of nuclei and nuclear matter.
The basic theorem given by Hohenberg and Kohn~\cite{PhysRev.136.B864}
guarantees the existence of the functional of one-body density.
The successful functionals for interacting fermi-particle systems are based on the Kohn--Sham scheme~\cite{PhysRev.140.A1133},
in which the density is represented in terms of ``single-particle'' (Kohn--Sham) orbitals in a potential that is a functional of density.
The problem should be solved self-consistently because the potential depends on the density and vice versa. 
The self-consistent solution corresponds to the density to minimize the energy density functional (EDF), namely the ground-state energy.

The static Kohn--Sham (mean-field) calculation
is not applicable to excited states.
In addition, it is difficult to describe states with large quantum fluctuations {using this calculation}, 
such as shape coexistence phenomena.
To describe these states, many theoretical approaches were proposed to incorporate correlations
beyond the mean-field with
the superposition of different configuration states,
such as the random phase approximation (RPA) \cite{RingSchuck},
and the generator coordinate method (GCM) \cite{PhysRev.108.311,RingSchuck}.
Among them, the~shell model calculation is one of the most successful methods to include beyond-mean-field correlations.
In the shell model method,
we first define a valence model space
to reduce the vast number of states in the configuration mixing to a manageable size.
The empirical Hamiltonian (effective interaction) is determined by fitting experimental spectra
with results of the 
configuration interaction calculation in the valence space.
In most cases, the single-particle energies are taken from the experimental data, and~the two-body matrix elements (TBMEs) are evaluated based on the $G$-matrix theory but are modified to reproduce the experimental values.
The USDB interaction~\cite{PhysRevC.74.034315} for the $sd$-shell model space and the SDPF-MU interaction~\cite{PhysRevC.86.051301} for the $sd$--$pf$ model space are known as examples of such empirical interactions.

In the present study, 
for a non-empirical description of nuclei
in a broad mass region,
we determine both the single-particle energies and the TBMEs of the shell model interactions employing a nuclear EDF.
In the non-relativistic density functionals, there are zero-range interactions of the Skyrme type~\cite{SKYRME1958615, PhysRevC.5.626}
and finite-range interactions of the Gogny type~\cite{PhysRevC.21.1568},
both of which are successful in the mean-field calculation to describe nuclear structure.
Several previous works have implemented the shell model method combined with a density functional.
The Skyrme-type effective interactions were tested for the shell model approach in Refs.~\cite{SAGAWA1985228, GOMEZ1993451}.
Some Skyrme interactions reasonably reproduce the experimental ground-state properties and low-lying excited spectra.
However, the~density-dependent interaction is treated in an approximate manner, in~which the density is not calculated with the shell model wave functions.
Reference~\cite{PhysRevC.98.044320} uses a Gogny-type effective interaction as the shell model effective interaction in a valence space.
The density-dependent interaction is self-consistently
treated,
in which the density is obtained with the shell model wave function.
It shows an ability to reproduce ground-state and excited-state properties in accuracy comparable with the empirical interactions
for nuclei in the mass region of $Z\sim 8$.
Since the DFT is designed to describe the nuclear properties in a broad mass region, 
it is important to investigate the ability of the shell model calculation using the density functional for the heavier mass~region.

In this paper, we adopt the hybrid approach following Ref.~\cite{PhysRevC.98.044320}
combining the Gogny-type density functionals with the shell model calculation, and investigate O, Ne, Mg, and~Ca isotopes to examine its capability for heavier nuclei. 
We compare our results with the experimental data, those obtained with the USDB interaction for O, Ne, and~Mg isotopes, and~with those with the SDPF-MU in Ca~isotopes.

\section{Theoretical framework}

A shell model Hamiltonian without assuming an inert core consists of the one-body and two-body interactions as
\begin{eqnarray}
  H &=& \sum\limits_{i,j} {t_{ij}a_{i}^{\dagger}a_{j}} + \frac{1}{4} \sum\limits_{i,j,k,l} {v_{ijkl} a_{i}^{\dagger}a_{j}^{\dagger}a_{l}a_{k}}, 
  \label{eq:nocoreH}
\end{eqnarray}
where $a^{\dagger}_i$ and $a_i$ are the creation and annihilation operators of the single-particle state $i$, and~
$t_{ij}$ and $v_{ijkl}$ are the antisymmetrized matrix elements of one-body and two-body terms, respectively.
In the no-core case, the~one-body matrix element is the kinetic energy 
and the two-body matrix element is given by
an interaction potential between two nucleons
which is density-dependent in the present case.
When we solve the Schr\"{o}dinger equation in the conventional shell model calculations, we assume a frozen inert core and the active particles are present only in the valence space.
The shell model Hamiltonian in the valence space is given by the following equation:
\begin{eqnarray}
  H_{\rm{SM}} &=&
  T_{\rm{SPE}} + \frac{1}{4} \sum\limits_{i,j,k,l \in \rm{Valence} } {v_{ijkl} a_{i}^{\dagger}a_{j}^{\dagger}a_{l}a_{k}}.
    \label{eq:smH}
\end{eqnarray}
The second term in the right-hand side of Equation~\eqref{eq:smH}
is the interaction between two nucleons in the valence space.
The first term, $T_{\rm{SPE}}$, denotes the one-body term given by
the normal ordering with respect to the inert core.
The single-particle energy of the orbit $i$ is evaluated as a sum of the kinetic energy and the two-body interaction between a nucleon in the core and one in the valence space as 
\begin{eqnarray}
  T_{\rm{SPE}} &=& \sum\limits_{i \in \rm{Valance}} { (t_{ii} + \sum\limits_{j \in \rm{Core}} v_{ijij}) a_{i}^{\dagger}a_{i}}
  . \notag
\end{eqnarray}
Here, 
 off-diagonal one-body matrix elements are omitted since we take the $0\hbar\omega$ model space in the present study. 
We further assume that they can be approximated
by the spherical harmonic-oscillator~basis.

In conventional shell model studies, the~single-particle energies and the two-body matrix elements (TBMEs) are phenomenologically determined to reproduce the experimental data.
In the present work, we adopt Gogny-type density functionals~\cite{PhysRevC.21.1568} as an effective interaction in shell model calculations.
The Gogny potential between nucleons 1 and 2 based on the functional is written as 
\begin{eqnarray}
  V{(\bm{r}_{1}, \bm{r}_{2})} &=& \sum\limits_{i = 1, 2} e^{-(\bm{r}_{1}^{2} - \bm{r}_{2}^{2}) / \mu_{i}^{2}} \left( W_{i} - B_{i} P_{\sigma} - H_{i} P_{\tau} - M_{i}P_{\sigma \tau} \right) \notag \\
         &+& i W_{\rm{LS}} \left( \sigma_{1} + \sigma_{2} \right) \cdot \bm{k'} \times \delta \left( \bm{r}_{1} - \bm{r}_{2} \right) \bm{k} \notag \\
         &+& t_{3} \left( 1 + x_{0} P_{\sigma} \right) \delta \left( \bm{r}_{1} - \bm{r}_{2} \right) \left[ {\rho \left( \frac{\bm{r}_{1} + \bm{r}_{2}}{2} \right)} \right]^{\alpha},
         \label{eq:d1s}
\end{eqnarray}
where the positions of the two interacting nucleons are $\bm{r}_{1}$ and $\bm{r}_{2}$,
and $\bm{k}$ and $\bm{k'}$ are the relative wave vectors.
The quantities $\mu_{i}$, $W_i$, $B_i$, $H_i$, $M_i$ ($i=1,2)$,
$W_{LS}$, $t_{3}$, $x_{0}$, and~$\alpha$ are parameters of Gogny functionals.
$P_{\sigma}$ and $P_{\tau}$ are spin- and isospin-exchange operators.
The first term corresponds to
a central potential, the~second term to a spin-orbit potential,
and the third term is a density-dependent potential.
This central force is a two-range Gaussian potential depending on spin and isospin with the widths $\mu_{1}$ and $\mu_{2}$.
The density-dependent force is widely used in nuclear DFT
that is important for the nuclear saturation
and incompressibility.
In the present study, we evaluate the matrix element $V_{ijkl}$ in Equation~(\ref{eq:nocoreH}) using this functional with the harmonic-oscillator single-particle wave functions and perform shell model calculations to obtain the ground-state and low-lying excited states.
The harmonic-oscillator frequency is determined by an empirical formula $\hbar \omega = 45A^{-1/3} - 25A^{-2/3}$ MeV~\cite{BLOMQVIST1968545}
to be in good agreement with the charge radii of
spherical~nuclei.

To compute the TBMEs of the Gogny interaction,
the nuclear density $\rho(\bm{r})$ in \mbox{Equation~(\ref{eq:d1s})}
is required.
In the present study, it is determined by performing the shell model calculations iteratively in a similar way to self-consistent mean-field methods. 
We start the Woods--Saxon density distribution as an initial nuclear density and compute the TBMEs of the Gogny interaction
of Equation~(\ref{eq:d1s}).
In the subsequent steps, we use a renewed ground-state density given by the shell model wave function and perform the shell model calculations iteratively. 
This iterative procedure continues until the energy converges sufficiently.
The Coulomb interaction is not considered in the shell model calculations for simplicity,
but taken into account for the ground-state energy as in Equation~\eqref{eq:Egs}.

The D1S parameter set is commonly used and successful
in the Gogny-type EDFs.
It was designed to fit experimental values of finite nuclei and reproduce the properties of nuclear matter.
In this work,
we use the Gogny D1S~\cite{BERGER1991365} in the shell model calculations and investigate $sd$-shell nuclei and Ca isotopes.
We perform the shell model calculations using the KSHELL code~\cite{SHIMIZU2019372}.

\vspace{-5pt}\begin{figure}[htbp]
  \centering
  \begin{minipage}[c]{0.45\columnwidth}
    \centering
    \includegraphics[width=\columnwidth]{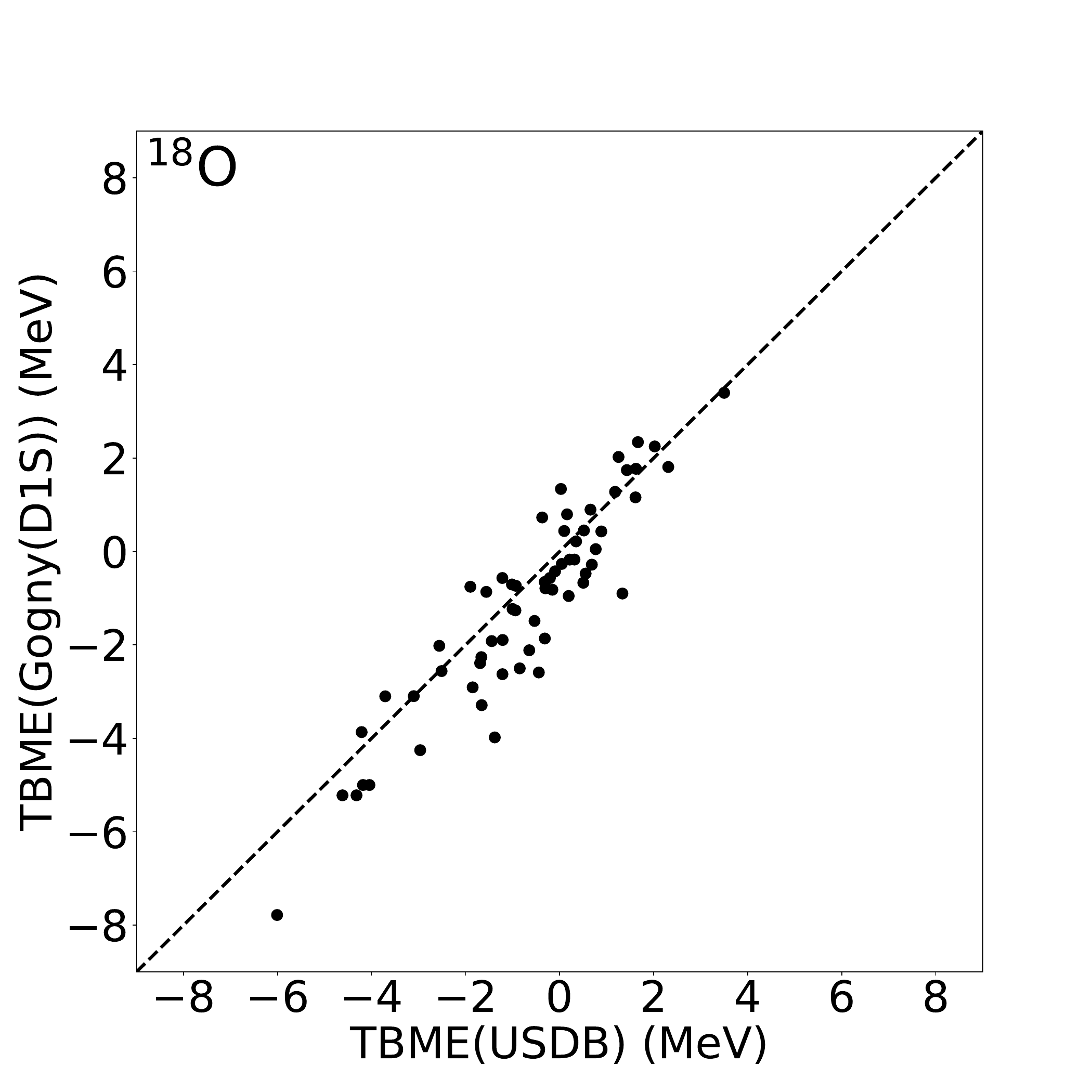} 
    \subcaption{}
    \label{fig:o16tbme}
  \end{minipage}
  \hspace{0.04\columnwidth}
  \begin{minipage}[c]{0.45\columnwidth}
    \centering
    \includegraphics[width=1.0\linewidth]{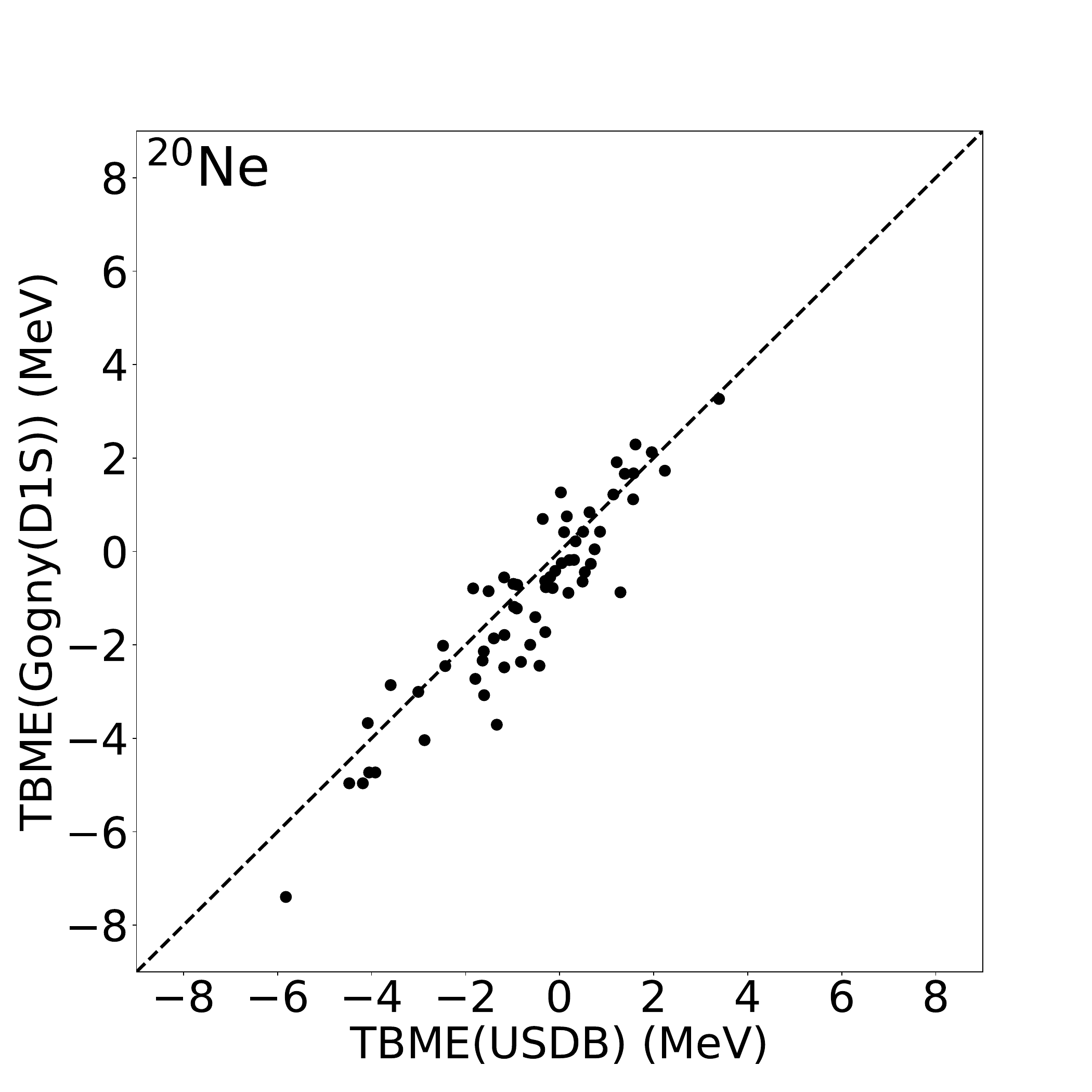}
    \subcaption{}
    \label{fig:ne20tbme}
  \end{minipage}
  \begin{minipage}[c]{0.45\columnwidth}
    \centering
    \includegraphics[width=1.0\linewidth]{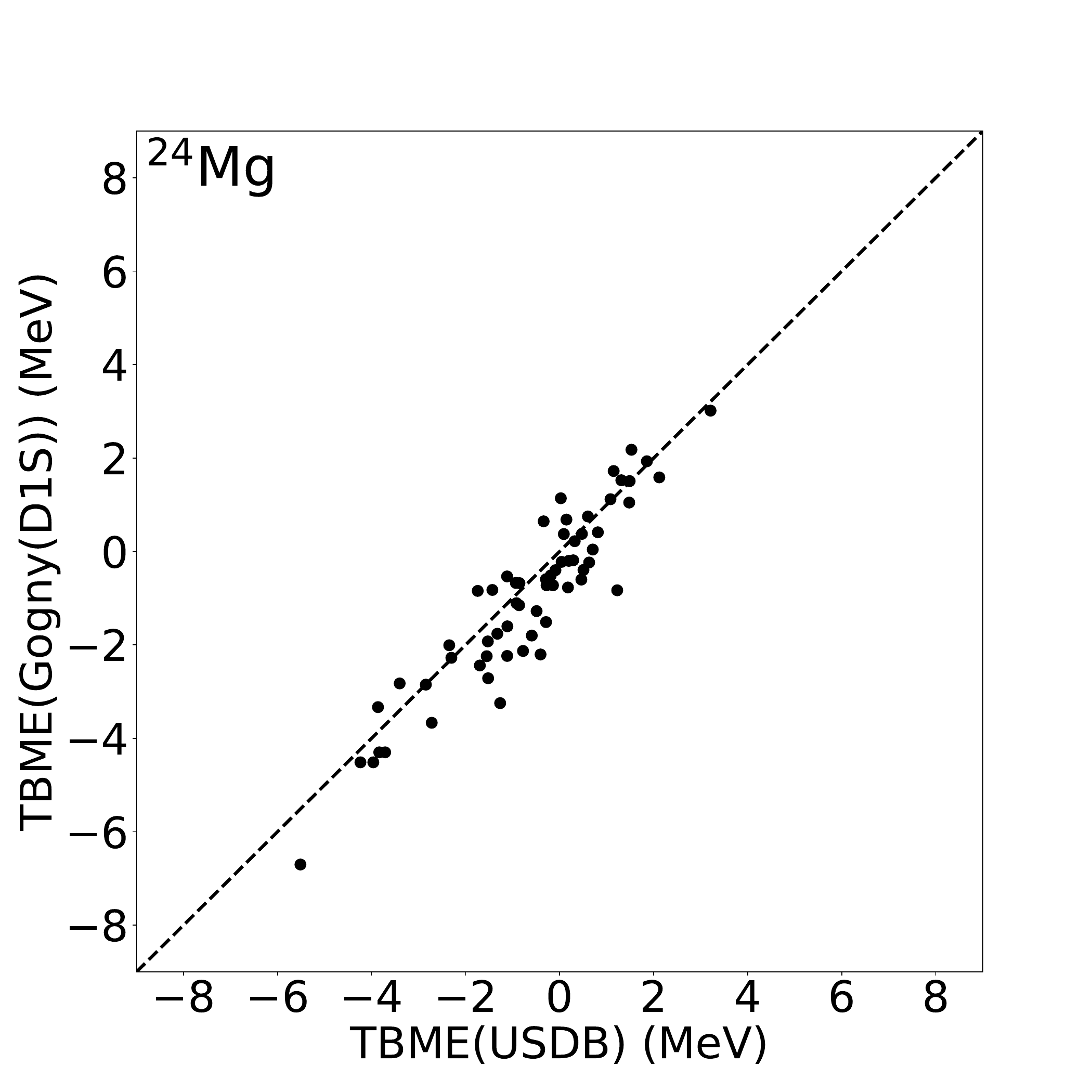}
    \subcaption{}
    \label{fig:mg24tbme}
  \end{minipage}
  \hspace{0.04\columnwidth}
  \begin{minipage}[c]{0.45\columnwidth}
    \centering
    \includegraphics[width=1.0\linewidth]{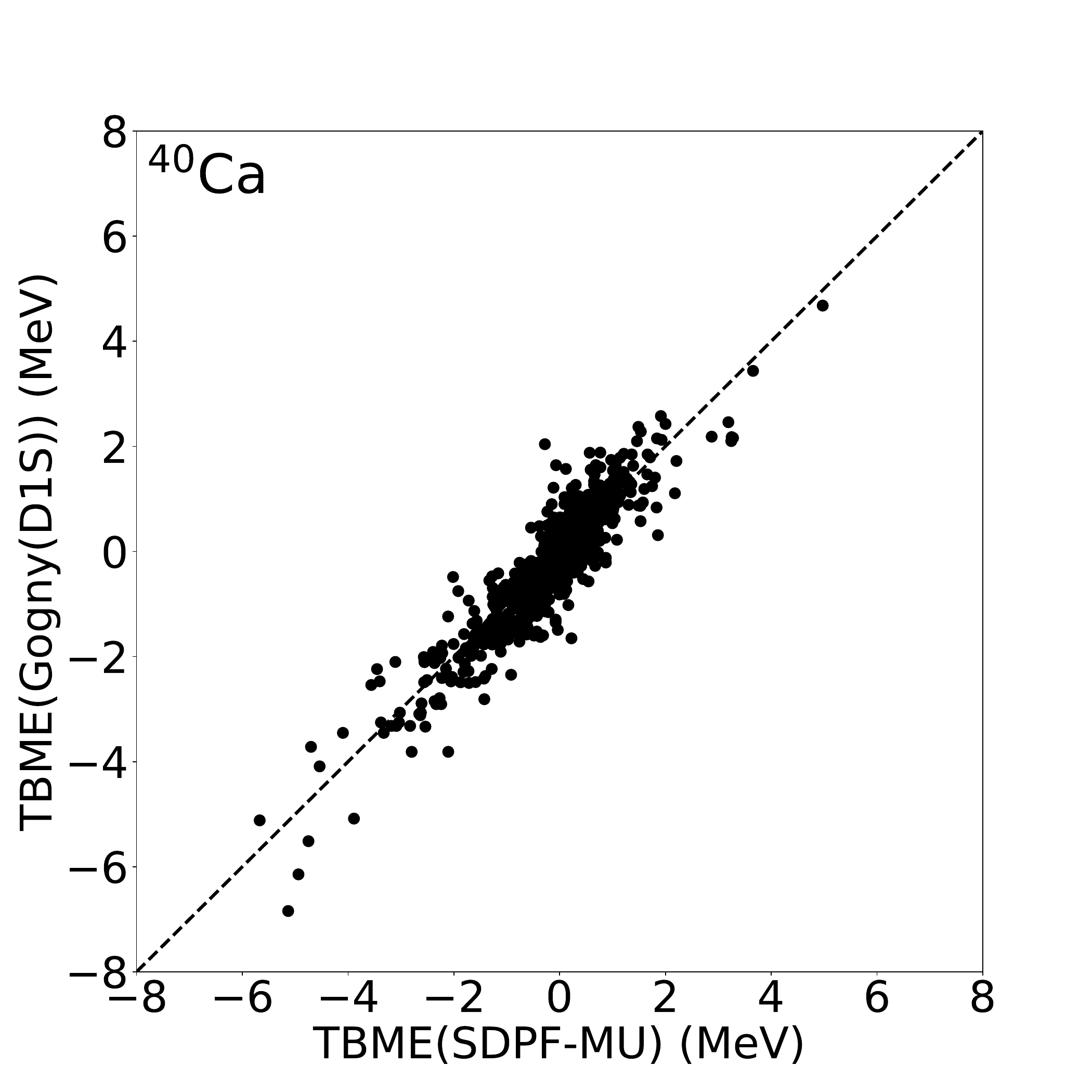}
    \subcaption{}
    \label{fig:ca40tbme}
  \end{minipage}
  \caption{
    Comparison of the two-body matrix elements (TBMEs) given by the Gogny-D1S functional and those of the empirical interactions (USDB, SDPF-MU).
    Figures (\textbf{a}--\textbf{c}) show the TBMEs for $^{18}$O, $^{20}$Ne, and~$^{24}$Mg, respectively, for~the $sd$ shell. Figure (\textbf{d}) shows the TBMEs of $^{40}$Ca in the $sd$--$pf$ shell.
    The shell model calculations for Ca isotopes in the following adopt only the $pf$ shell as the valence space.
    The dashed lines correspond to the diagonal lines, on which the TBMEs of the Gogny-D1S and the empirical interactions are identical.
  }
  \label{fig:tbme}
\end{figure}

\section{Results and discussion}

We examined the validity of the present hybrid model by comparing
the results with those of the phenomenological shell model interactions and~experimental data.
The valence space was taken as $sd$ shell for $sd$-shell nuclei and $pf$ shell for Ca~isotopes. 

First, we compared the TBMEs of the Gogny-D1S interaction with those of the empirical shell model interactions.
We calculated the TBMEs with the harmonic-oscillator wave function for  $^{18}$O, $^{20}$Ne, $^{24}$Mg, and $^{40}$Ca using the Gogny-D1S interaction and compared them with the USDB and SDPF-MU interactions.
Figure~\ref{fig:tbme}a--d show the correlation between the TBMEs of the Gogny-D1S interaction and those of the empirical interactions.
The points are somewhat scattered but close to the diagonal line, and the Gogny-D1S TBMEs agree with the empirical ones roughly within 2 MeV.
This means that the Gogny-D1S TBMEs are close to those of USDB
in the $sd$ shell and of SDPF-MU in the $sd$--$pf$ shell.
The density-dependent term in {Gogny D1S} 
 only has the $T = 0$ channel because of $x_{0} = 1$ in Equation~(\ref{eq:d1s}).
Almost all the $T=0$ TBMEs of the central force and those of the density-dependent force have opposite signs,
leading to some cancellation in the magnitude of TBMEs.
The Gogny functional is written as the finite-range Gaussian functions and the high-momentum component is renormalized into the low-momentum space, which is advantageous for the application to the shell model calculations.
The USDB interaction is based on the $G$-matrix theory; however, the~matrix elements were fitted to reproduce the experimental values of $A = 16$--$40$ nuclei.
Note that the TBMEs of the USDB interaction have the mass dependence factor $(A/18)^{-0.3}$, which is well described by the Gogny ones. 
In Figure~\ref{fig:tbme}d,{ we can see that a~large portion} 
of the TBMEs of $^{40}$Ca are concentrated
in a range of $[-3,3]$ MeV.

\begin{figure}[h]
\includegraphics[width=15 cm]{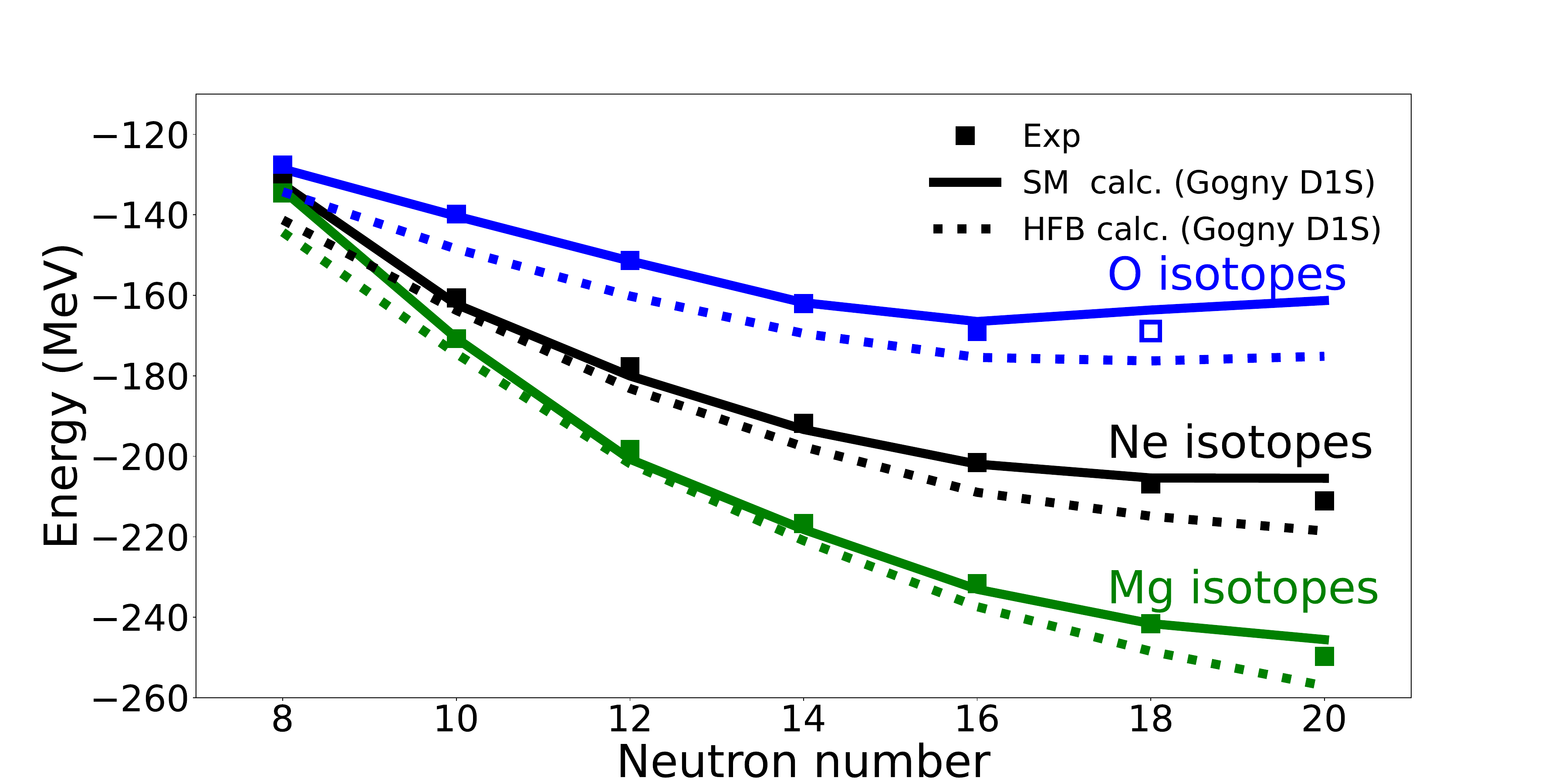}
\caption{
Ground-state energies of the O, Ne, and~Mg isotopes against the neutron number. The~solid line, dotted line, and~filled squares denote the energies provided by the present shell model calculations,
the HFB, and~the experiments~\cite{Wang2021}, respectively. 
$^{26}$O is unbound and
the open square denotes experimental data of 
the resonance state.
}
\label{fig:sdshellegs}
\end{figure}

\begin{figure}[h]
\includegraphics[width=15 cm]{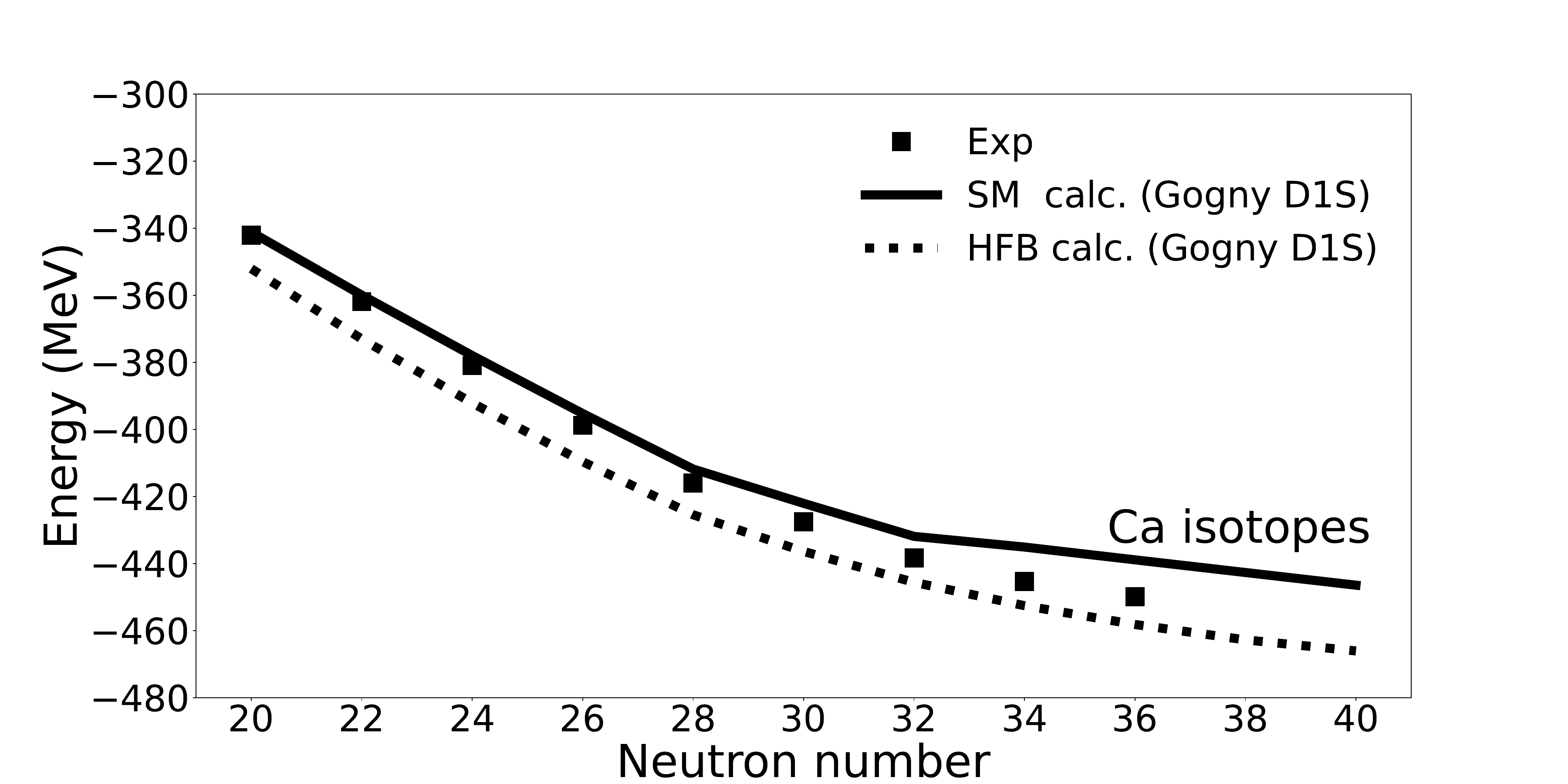}
\caption{
Ground-state energies of even-mass Ca isotopes. See the caption of Figure~\ref{fig:sdshellegs} for details.
\label{fig:caegs}}
\end{figure}

Figures~\ref{fig:sdshellegs} and \ref{fig:caegs} show the ground-state energies of the Gogny-D1S shell model calculation,
the mean-field calculations,
and the experimental data~\cite{Wang2021}
for even--even $sd$-shell nuclei and Ca isotopes.
The ground-state energies ($E_\textrm{g.s.}$) are given by
the sum of 
the shell model energy ($E_{\rm{SM}}$), the~energy of the inert core ($E_{\rm{Core}}$), the~Coulomb energy 
($E_{\rm{Coul.}}$),
and the correction associated with
the kinetic energy of the center-of-mass motion ($E_{\rm{CoM}}$) as
\begin{eqnarray}
  E_{\rm{g.s.}} = E_{\rm{SM}} + E_{\rm{Core}} + E_{\rm{Coul.}} - E_{\rm{CoM}} .
  \label{eq:Egs}
\end{eqnarray}
For simplicity, we assume the isospin symmetry and the Coulomb energy is estimated by an empirical formula \cite{PhysRevC.97.054321}.

The mean-field results shown in Figures~\ref{fig:sdshellegs} and \ref{fig:caegs} are obtained using the Hartree--Fock--Bogoliubov (HFB) method~\cite{RingSchuck}. 
We performed these mean-field calculations using the HFBTHO code~\cite{STOITSOV20131592}.
These HFB results of O, Ne, Mg, and~Ca isotopes overestimate the experimental binding energies.
The two-body center-of-mass correction removes a major part of this discrepancy, 
which will be discussed later. 
In the Ne and Mg isotopes, the~deviation of the HFB results increases as the neutron number increases away from the $N=8$ magic~number.

\vspace{-5pt}\begin{figure}[htbp]
  \centering
  \begin{minipage}[c]{0.45\columnwidth}
    \centering
    \includegraphics[width=\columnwidth]{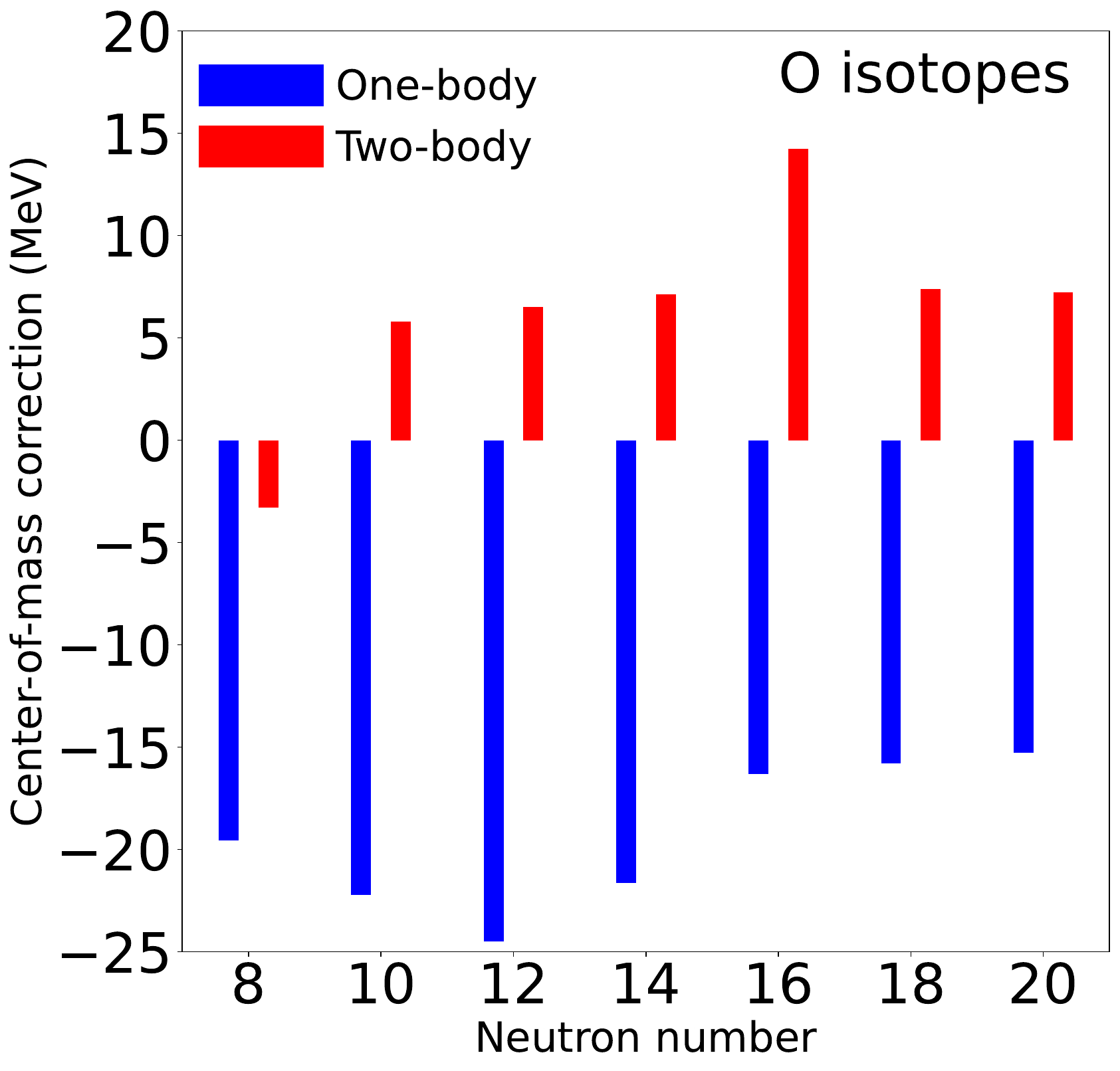} 
    \subcaption{}
  \end{minipage}
  \hspace{0.04\columnwidth}
  \begin{minipage}[c]{0.45\columnwidth}
    \centering
    \includegraphics[width=1.0\linewidth]{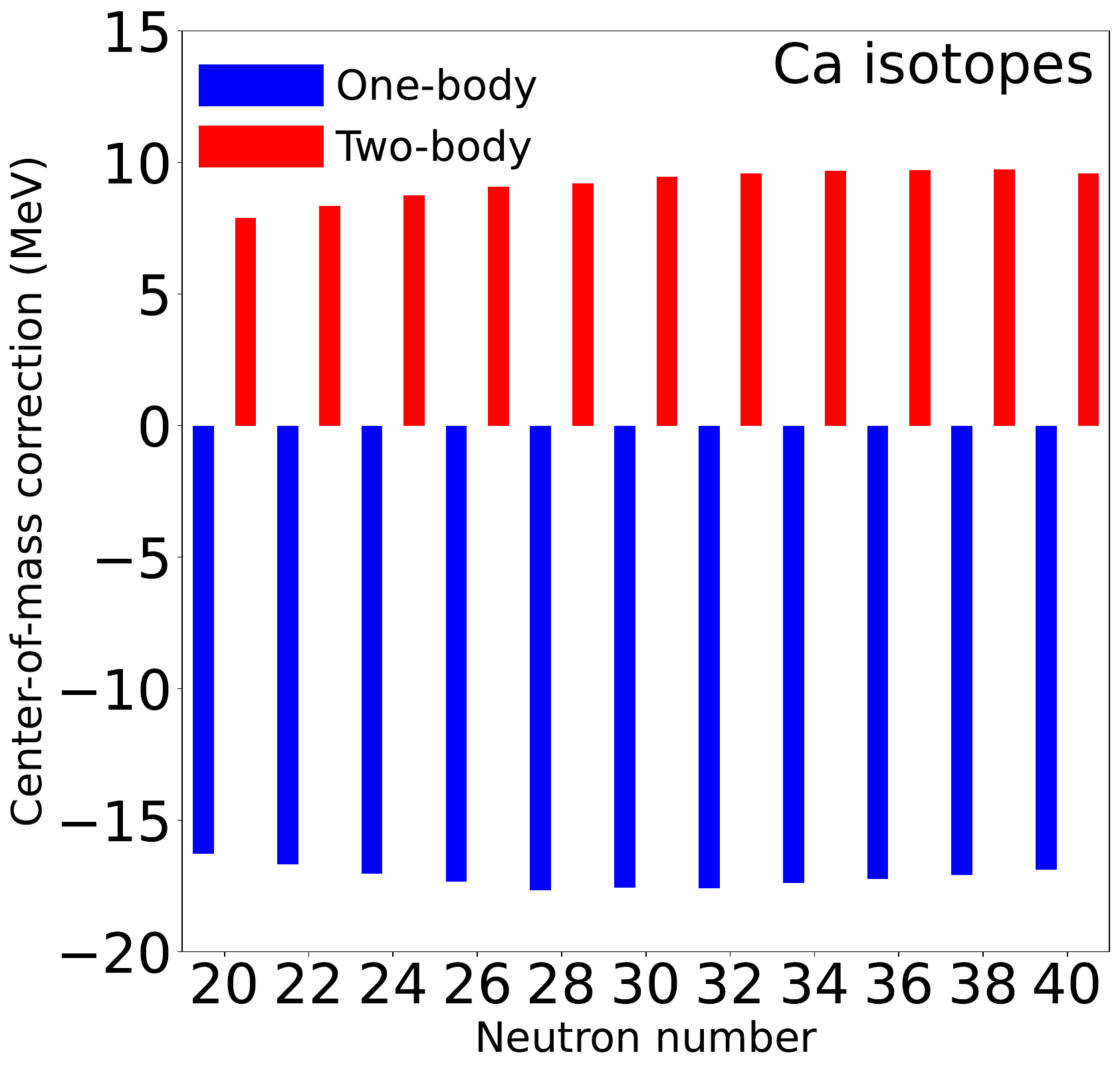}
    \subcaption{}
  \end{minipage}
  \caption{
    Contribution of the center-of-mass correction for the O isotopes (\textbf{a}) and Ca isotopes (\textbf{b})~against the neutron number.
    The blue (red) bars show the values of the one-body (two-body)~corrections.
  }
  \label{fig:com}
\end{figure}

In order to investigate the discrepancy between the HFBTHO results and the experimental values, we here discuss the center-of-mass correction since the HFBTHO code does not include the two-body part of this correction. 
The total energy of the nucleus should be evaluated by subtracting the kinetic energy of the center-of-mass motion.
The center-of-mass energy is given as
\begin{align}
E_{\rm{CoM}}
&=\frac{1}{2Am_{N}} \left( \sum_{ij}\left\langle {i} \right| {\bf p} \left| {j} \right\rangle a_{i}^{\dagger} a_{j} \right)^2
\nonumber \\
  &= \frac{1}{2Am_{N}} \sum_{ij}{\left\langle {i} \right|  {\bf p}^{2} \left| {j} \right\rangle  a_{i}^{\dagger} a_{j}} 
  + \frac{1}{2Am_{N}} \sum_{ijkl}
  {\left\langle {i} \right|  {\bf p} \left| k\right\rangle
  \cdot
  \left\langle {j} \right|  {\bf p} \left| l\right\rangle
  a_{i}^{\dagger} a_{j}^{\dagger} a_{l} a_{k}} ,
  \label{eq:com}
\end{align}
where $A$ corresponds to the mass number,
$m_{N}$ to the nucleon mass, and~${\bf p}$ to the nucleon's momentum.
The first and second terms of Equation~\eqref{eq:com} are often called the one-body and two-body center-of-mass corrections, respectively.
In the current hybrid approach with the $0\hbar\omega$ model space, the~contribution of the two-body correction vanishes, and~the total correction is estimated as $\frac34 \hbar \omega$ \cite{PhysRevC.98.044320}.
In the mean-field calculations including the HFB results with the HFBTHO code in Figures~\ref{fig:sdshellegs} and \ref{fig:caegs},
the two-body corrections exist in general but are neglected.
We compute the contribution of the two-body correction
in the HFB calculations using the HFBSPH code~\cite{HFBSPH}.
Since the spherical symmetry is assumed in this code, 
we show the results for the O and Ca isotopes
in Figure~\ref{fig:com}.
As we expect, the~total contribution of the corrections is always negative since the kinetic energy of the center-of-mass motion is positive. 
However, the~contribution of the two-body correction is positive in most cases, and~is around 8 MeV for the Ca isotopes.
Thus, the~ground-state energies given by the HFBTHO code in Figure~\ref{fig:caegs} are expected to be lifted up roughly by 8 MeV by introducing the two-body center-of-mass correction and to approach the experimental values.

The shell model results successfully reproduce the experimental values of the $sd$-shell nuclei up to $N = 16$ and $^{40\text{--
}48}$Ca.
In the neutron-rich region of $^{50\text{--}60}$Ca,
the binding energy is underestimated and the deviation
from the experimental data increases with the neutron number.
Our shell model results tend to underestimate the experimental binding energies.
There may be two possible improvements on the current model:
To enlarge the model space and to optimize
the harmonic-oscillator frequency $\hbar \omega$.
The $0\hbar\omega$ valence model space is adopted
in the present paper, but~
the cross-shell excitations could be important near
the end of the shell.
Although we use the empirical formula
\cite{BLOMQVIST1968545} for the harmonic-oscillator frequency $\omega$,
it may not be
adequate for ground-state energy in the neutron-rich regions.
Nevertheless,
$^{24}$O is the last bound nucleus at the the neutron dripline, which is correctly predicted by the Gogny shell model results, while the HFB result fails to reproduce it. 

\begin{figure}[h]
\includegraphics[width=16 cm]{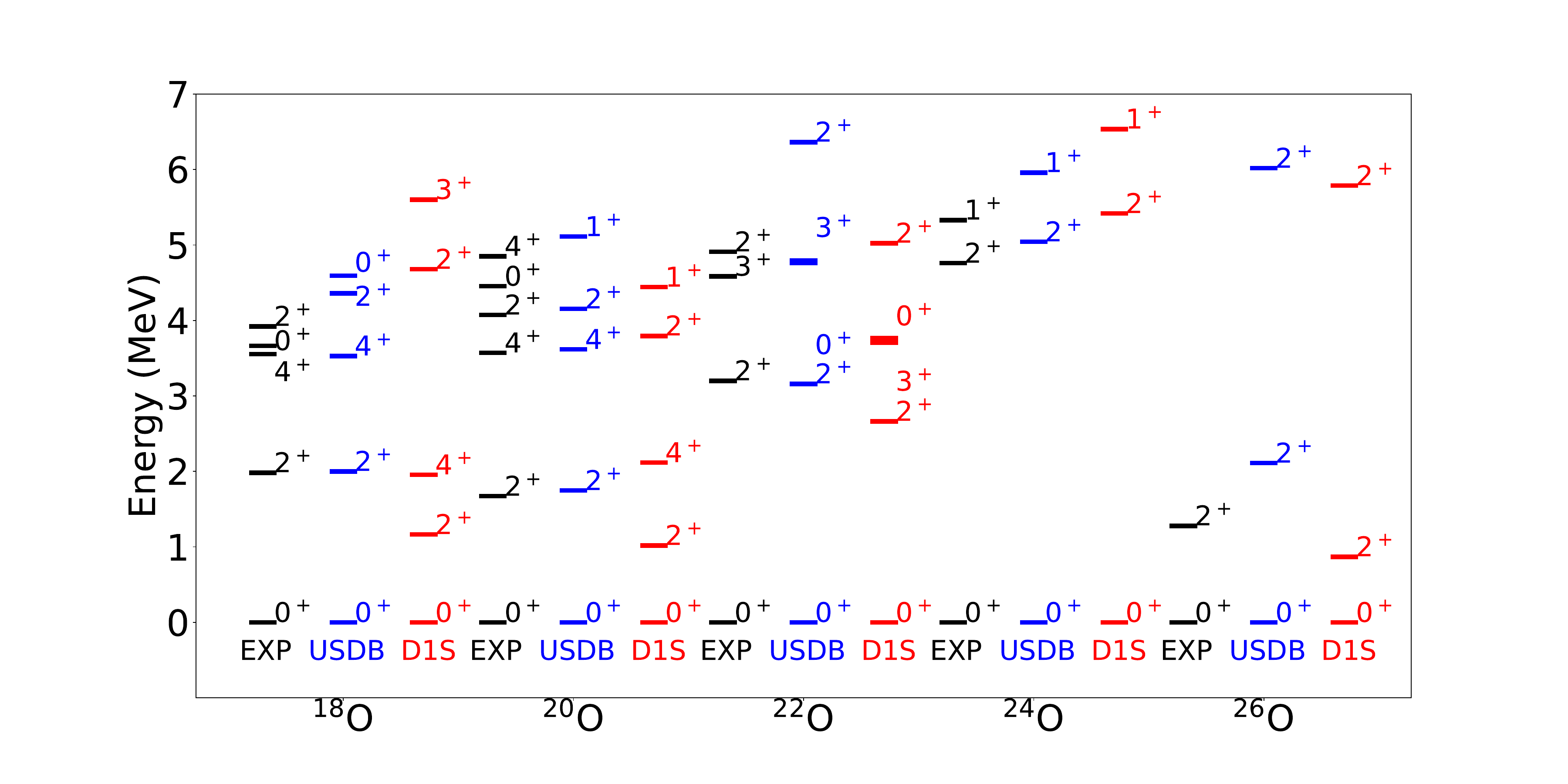}
\caption{Calculated excitation spectra with Gogny D1S and
USDB,
compared with the experimental data \cite{AUDI1995409, TILLEY1998249, BASUNIA201569, BASUNIA20223, BASUNIA20161, SHAMSUZZOHABASUNIA20131189, BASUNIA20241, CHEN20161, CHEN20231, WU20001, CHEN20221, CHEN20191}
for even-even O isotopes.
\label{fig:ospectra}}
\end{figure}

\begin{figure}[h]
\includegraphics[width=16 cm]{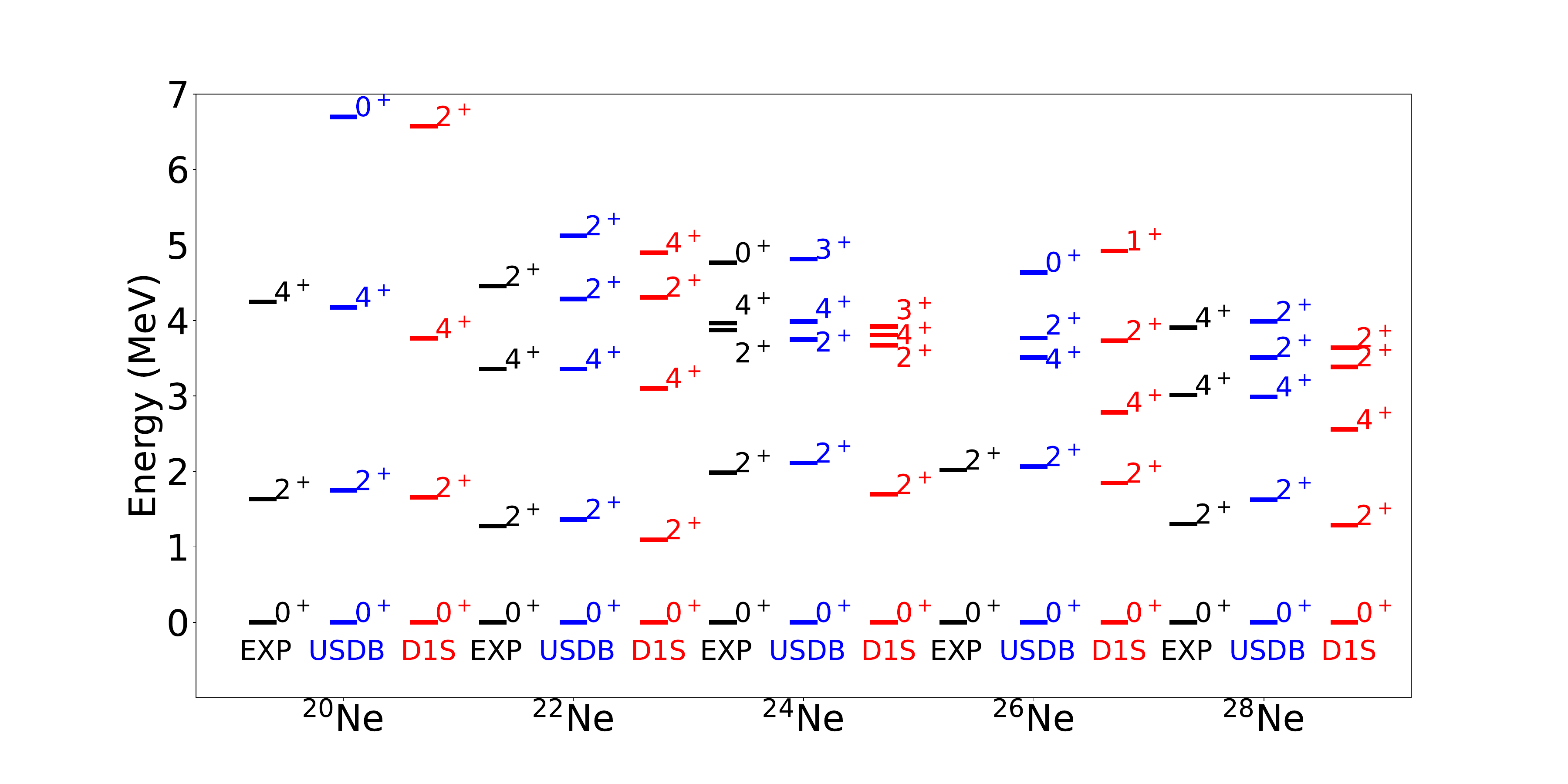}
\caption{Same as Fig. \ref{fig:ospectra}, but for even-even Ne isotopes.
\label{fig:nespectra}}
\end{figure}

\begin{figure}[h]
\includegraphics[width=16 cm]{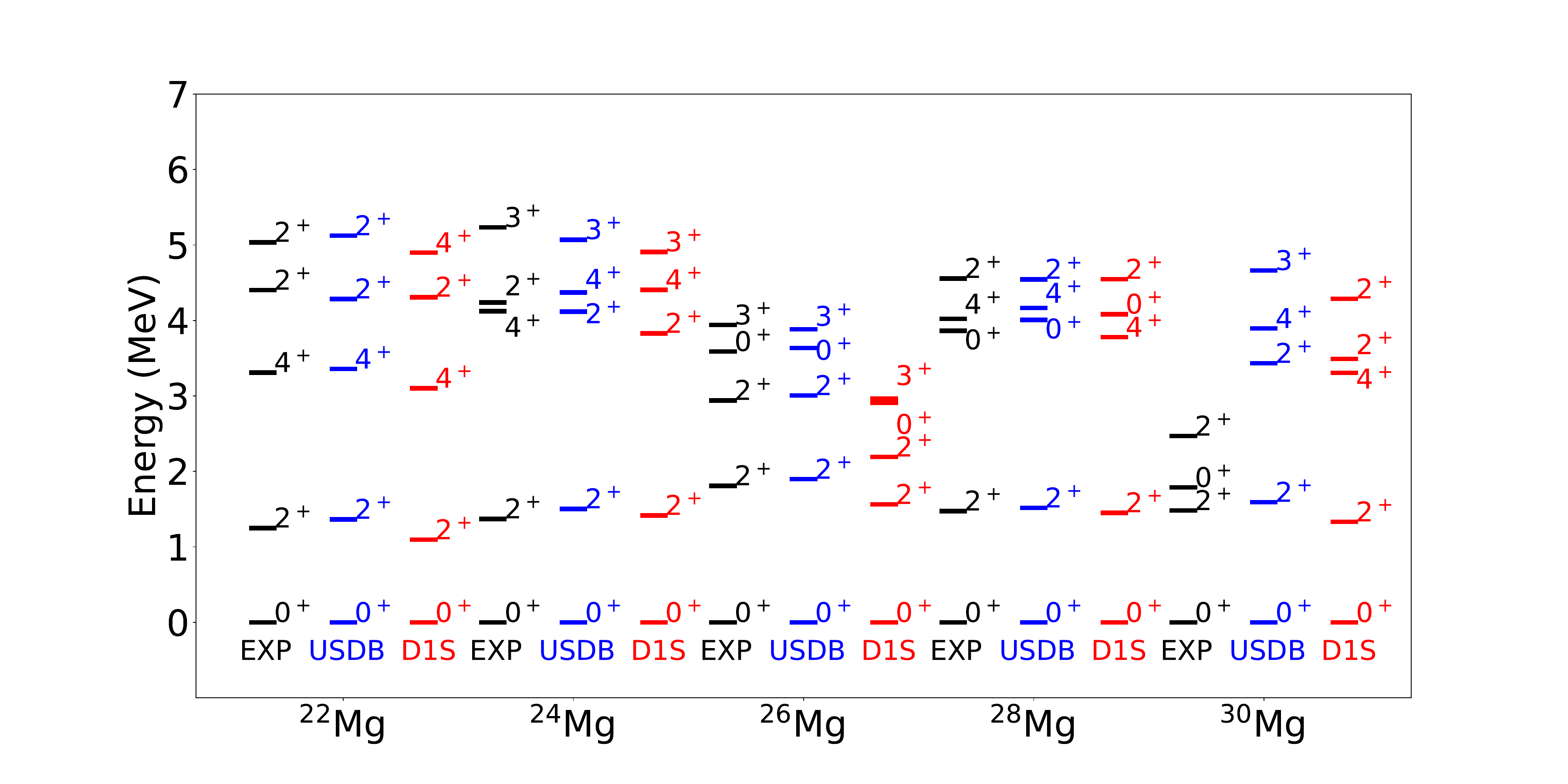}
\caption{Same as Fig. \ref{fig:ospectra}, but for even-even Mg isotopes.
\label{fig:mgspectra}}
\end{figure}

\begin{figure}[h]
\includegraphics[width=16 cm]{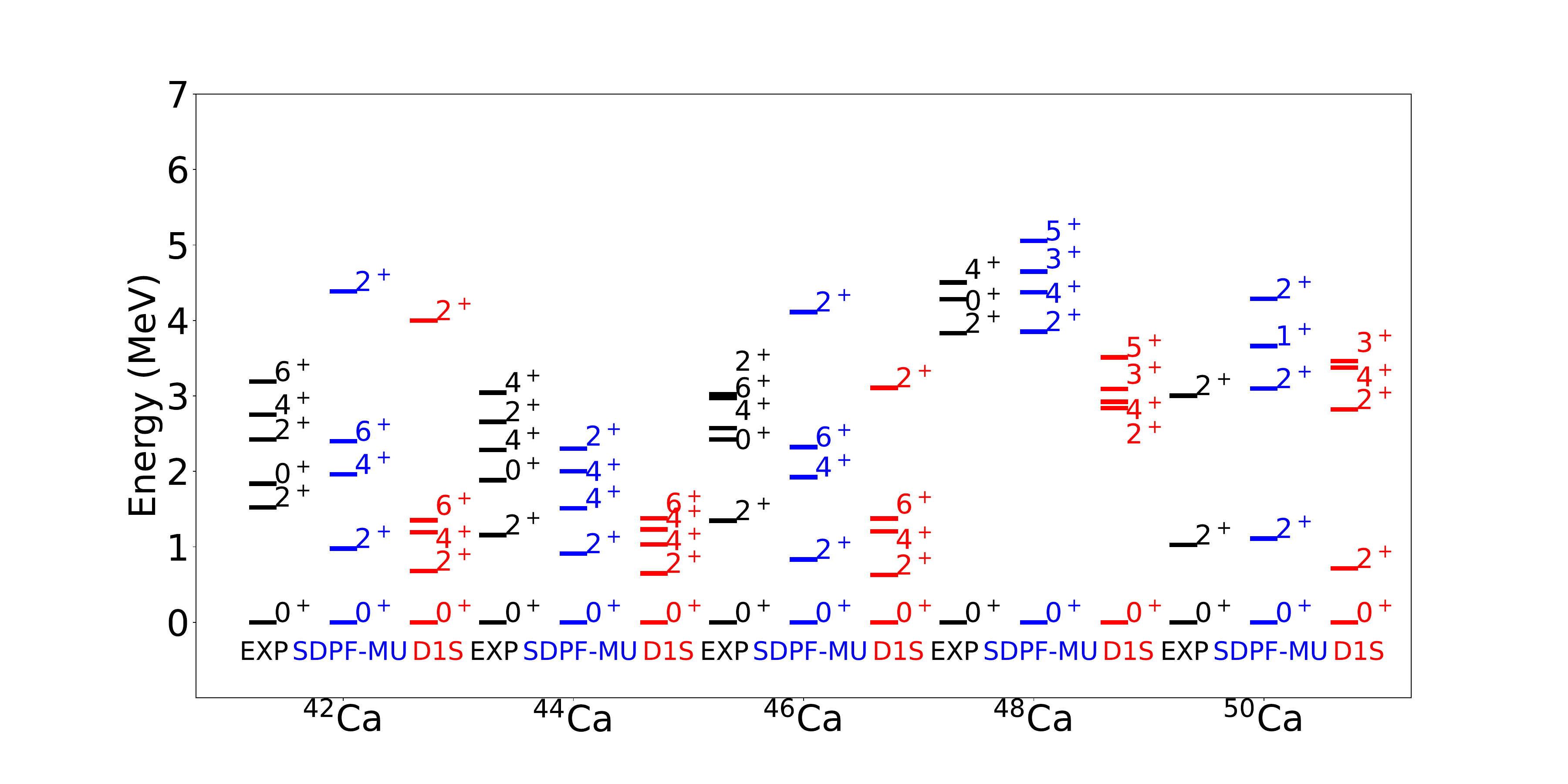}
\caption{Same as Fig. \ref{fig:ospectra}, but for even-even Ca isotopes.
The valence space is the \textit{pf} shell and
the phenomenological interaction is SDPF-MU instead of USDB.
\label{fig:caspectra}}
\end{figure}

Figures~\ref{fig:ospectra}--\ref{fig:caspectra} show the energy spectra for the Gogny-D1S
shell model calculations with the experimental data~\cite{AUDI1995409, TILLEY1998249, BASUNIA201569, BASUNIA20223, BASUNIA20161, SHAMSUZZOHABASUNIA20131189, BASUNIA20241, CHEN20161, CHEN20231, WU20001, CHEN20221, CHEN20191},
together with the USDB calculations
(O, Ne, and~Mg isotopes) and
with the SDPF-MU calculations (Ca isotopes).
Figure~\ref{fig:ospectra} demonstrates that the Gogny-D1S result matches the excitation energies, spin, and~parity of the experimental data of $^{24}$O and $^{26}$O.
The large $2^+$ excitation energy of $^{24}$O indicates the $N=16$ sub-shell gap, which is reproduced by both the USDB interaction and the Gogny-D1S interaction. 
The shell model result with Gogny D1S underestimates the experimental excitation energies of $^{18}$O, $^{20}$O, and~$^{22}$O, possibly due to the assumption of the inert core $^{16}$O.
In Figures~\ref{fig:nespectra} and \ref{fig:mgspectra},
the Gogny-D1S results for the Ne and Mg isotopes agree with the experimental data,
with the accuracy comparable to the USDB.
These results contrast with the O isotopes.
For nuclei with $N=10$ and 12, for~instance,
the $2^+_1$ energies are well reproduced for $^{20,22}$Ne
but significantly underestimated for $^{18,20}$O.
There is a significant drop in the excitation energy of
the $2^+_2$ state in $^{26}$Mg,
which is reproduced in the calculation.
In summary, 
the low-lying excited states for these nuclei
are well described
using the $sd$-shell model space. 

Figure~\ref{fig:caspectra} shows that the Gogny-D1S shell model calculations in the $pf$ shell
with an inert core $^{40}$Ca 
reasonably agree with
the experimental data in $^{48}$Ca and $^{50}$Ca.
The large $2^+$ excitation energy of $^{48}$Ca indicates the doubly magic nature,
although the Gogny-D1S shell model calculation underestimates
its magnitude.
The results of the $2^+_1$ energies are systematically smaller
than the experimental data in the Ca isotopes,
which is analogous to the case for the O isotopes.
This may also be due to the core excitations neglected in the
present calculations.

\begin{figure}[h]
\includegraphics[width=11 cm]{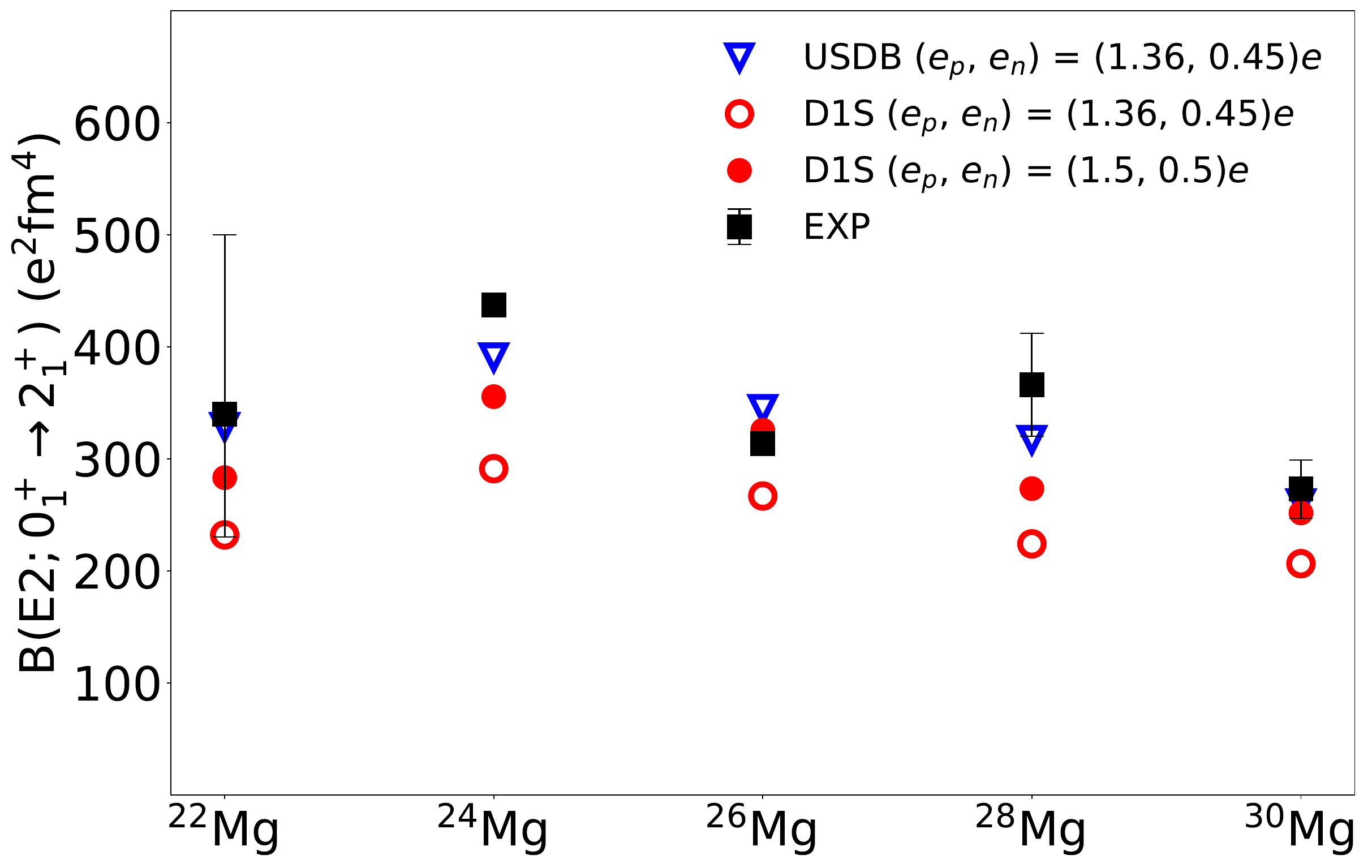}
\caption{$B(E2;0_{1}^{+} \rightarrow 2_{1}^{+})$ values of Mg isotopes against the neutron number.
These values are compared with USDB, Gogny D1S, and~experimental data~\cite{PRITYCHENKO20161}.
The shell model calculations were performed with the $sd$-shell model space.
The blue triangles, the~red circles, and~the black squares denote the USDB results, the~Gogny-D1S results, and~the experimental values, respectively.
All USDB results are given by the effective charges $(e_{p}, e_{n}) = (1.36,  0.45)e$.
The Gogny-D1S results are shown with the two different effective charges, $(e_{p}, e_{n}) = (1.36,  0.45)e$ (open red circles) and $(e_{p}, e_{n}) = (1.5,  0.5)e$  (filled red circles).
\label{fig:be2mg}}
\end{figure}   

To confirm the validity of the present approach further, we computed the $B(E2; 0^+_1 \rightarrow 2^+_1)$ transition probabilities of the Mg isotopes. 
Figure~\ref{fig:be2mg} shows the shell model results of the Mg isotopes using the Gogny-D1S and USDB interactions with the $sd$-shell model space compared to the experimental data~\cite{PRITYCHENKO20161}.
Note that the results of the Ne isotopes have been shown in Ref.~\cite{PhysRevC.98.044320}.
The Gogny-D1S results are systematically smaller than the USDB results with the same effective charges, $(e_{p},e_{n}) = (1.36, 0.45)e$, which were determined by the chi-squared fit for the USDB interaction, and~tend to underestimate the experimental $B(E2)$ values.
The agreement is improved to some extent by introducing typical effective charges, $(e_{p},e_{n}) = (1.5, 0.5)e$. 
Enlarging the model space may improve the agreement, which will be investigated as future work.

\section{Summary}

We have investigated $sd$-shell nuclei and Ca isotopes using a hybrid approach of the shell model and Gogny-type density functionals.
The model reproduces the experimental results of energy spectra with an accuracy comparable to the existing empirical interactions in O, Ne, Mg, and~Ca isotopes.
It is demonstrated that the $B(E2)$ transition probabilities of the Mg isotopes given by the present approach show reasonable agreement with the experimental values using the standard effective charges as well as the conventional shell model study.
While the empirical interaction consisting of single-particle energies and TBMEs must be prepared by fitting to experimental data in each valence model space,
in the present hybrid model, the~interaction is constructed using
an EDF determined to reproduce the ground-state properties
of several nuclei in the nuclear chart and the nuclear matter.
We expect this model to be applicable not only to O, Ne, Mg, and~Ca isotopes but also to nuclei in a broad mass region.
The model could be a powerful tool for describing heavy unknown nuclei
including correlations beyond mean-field with good accuracy.
On the other hand, the~present model underestimates
the ground-state energies in the neutron-rich region and
the excitation energies near the closed-shell configurations.
It may indicate the importance of the core excitations.
In future work, we plan to include further configuration mixing by extending the valence model space, 
and to employ other types of EDFs.

For future applications of the current model to the heavier-mass nuclei, the~shell model dimension often becomes too large to be treated. In~such a case, we plan to employ numerical methods such as the Monte Carlo shell model~\cite{mcsm_review} and quasi-particle vacua shell model~\cite{PhysRevC.103.014312}, which enable us to perform shell model studies in heavy-mass deformed nuclei~\cite{PhysRevC.108.L021302}.\vspace{6pt}

\acknowledgments{
We acknowledge Hitoshi Nakada for fruitful discussions and Daisuke Abe for his HFB code.
This work is supported by JST ERATO Grant No. JPMJER2304,
JST SPRING Grant No. JPMJSP2124, Japan,
and  KAKENHI Grant Nos. JP24H00239 and  JP23K25864.
N.S. acknowledges the support of the ``Program for promoting research on
the supercomputer Fugaku'', MEXT, Japan (JPMXP1020230411).
}

\nocite{*}


\begin{thebibliography}{999}

\bibitem[Hohenberg and Kohn(1964)]{PhysRev.136.B864}
Hohenberg, P.; Kohn, W.
\newblock Inhomogeneous Electron Gas.
\newblock {\em Phys. Rev.} {\bf 1964}, {\em 136},~B864--B871.
\newblock {\url{https://doi.org/10.1103/PhysRev.136.B864}}.

\bibitem[Kohn and Sham(1965)]{PhysRev.140.A1133}
Kohn, W.; Sham, L.J.
\newblock Self-Consistent Equations Including Exchange and Correlation Effects.
\newblock {\em Phys. Rev.} {\bf 1965}, {\em 140},~A1133--A1138.
\newblock {\url{https://doi.org/10.1103/PhysRev.140.A1133}}.

\bibitem[Ring and Schuck(1980)]{RingSchuck}
Ring, P.; Schuck, P.
\newblock {\em The Nuclear Many-Body Problem}; Springer: Berlin/Heidelberg, Germany, 
1980.

\bibitem[Griffin and Wheeler(1957)]{PhysRev.108.311}
Griffin, J.J.; Wheeler, J.A.
\newblock Collective Motions in Nuclei by the Method of Generator Coordinates.
\newblock {\em Phys. Rev.} {\bf 1957}, {\em 108},~311--327.
\newblock {\url{https://doi.org/10.1103/PhysRev.108.311}}.

\bibitem[Brown and Richter(2006)]{PhysRevC.74.034315}
Brown, B.A.; Richter, W.A. New  \mbox{``USD''} Hamiltonians for the \emph{sd} shell.
\newblock {\em Phys. Rev. C} {\bf 2006}, {\em 74},~034315.
\newblock {\url{https://doi.org/10.1103/PhysRevC.74.034315}}.

\bibitem[Utsuno et~al.(2012)Utsuno, Otsuka, Brown, Honma, Mizusaki, and
  Shimizu]{PhysRevC.86.051301}
Utsuno, Y.; Otsuka, T.; Brown, B.A.; Honma, M.; Mizusaki, T.; Shimizu, N. Shape transitions in exotic Si and S isotopes and tensor-force-driven Jahn-Teller effect.
\newblock {\em Phys. Rev. C} {\bf 2012}, {\em 86},~051301.
\newblock {\url{https://doi.org/10.1103/PhysRevC.86.051301}}.

\bibitem[Skyrme(1958)]{SKYRME1958615}
Skyrme, T.
\newblock The effective nuclear potential.
\newblock {\em Nucl. Phys.} {\bf 1958}, {\em 9},~615--634.
\newblock {\url{https://doi.org/https://doi.org/10.1016/0029-5582(58)90345-6}}.

\bibitem[Vautherin and Brink(1972)]{PhysRevC.5.626}
Vautherin, D.; Brink, D.M.
\newblock Hartree-Fock Calculations with Skyrme's Interaction. I. Spherical
  Nuclei.
\newblock {\em Phys. Rev. C} {\bf 1972}, {\em 5},~626--647.
\newblock {\url{https://doi.org/10.1103/PhysRevC.5.626}}.

\bibitem[Decharg\'e and Gogny(1980)]{PhysRevC.21.1568}
Decharg\'e, J.; Gogny, D. Hartree-Fock-Bogolyubov calculations with the \emph{D}1 effective interaction on spherical nuclei.
\newblock {\em Phys. Rev. C} {\bf 1980}, {\em 21},~1568--1593.
\newblock {\url{https://doi.org/10.1103/PhysRevC.21.1568}}.

\bibitem[Sagawa et~al.(1985)Sagawa, Brown, and Scholten]{SAGAWA1985228}
Sagawa, H.; Brown, B.; Scholten, O.
\newblock Shell-model calculations with a skyrme-type effective interaction.
\newblock {\em Phys. Lett. B} {\bf 1985}, {\em 159},~228--232.
\newblock {\url{https://doi.org/https://doi.org/10.1016/0370-2693(85)90240-0}}.

\bibitem[Gómez et~al.(1993)Gómez, Cerdán, and Prieto]{GOMEZ1993451}
Gómez, J.; Cerdán, J.; Prieto, C.
\newblock Shell-model description of nuclei with 4 $\geq$ A $\geq$ 16 using
  Skyrme forces.
\newblock {\em Nucl. Phys. A} {\bf 1993}, {\em 551},~451--472.
\newblock {\url{https://doi.org/https://doi.org/10.1016/0375-9474(93)90457-9}}.

\bibitem[Jiang et~al.(2018)Jiang, Hu, Sun, and Xu]{PhysRevC.98.044320}
Jiang, W.G.; Hu, B.S.; Sun, Z.H.; Xu, F.R. Gogny-force-derived effective shell-model Hamiltonian.
\newblock {\em Phys. Rev. C} {\bf 2018}, {\em 98},~044320.
\newblock {\url{https://doi.org/10.1103/PhysRevC.98.044320}}.

\bibitem[Blomqvist and Molinari(1968)]{BLOMQVIST1968545}
Blomqvist, J.; Molinari, A. Collective 0$^{-}$ vibrations in even spherical nuclei with tensor forces.
\newblock {\em Nucl. Phys. A} {\bf 1968}, {\em 106},~545--569.
\newblock {\url{https://doi.org/10.1016/0375-9474(68)90515-0}}.

\bibitem[Berger et~al.(1991)Berger, Girod, and Gogny]{BERGER1991365}
Berger, J.; Girod, M.; Gogny, D. Time-dependent quantum collective dynamics applied to nuclear fission.
\newblock {\em Comput. Phys. Commun.} {\bf 1991}, {\em 63},~365--374.
\newblock {\url{https://doi.org/10.1016/0010-4655(91)90263-K}}.

\bibitem[Shimizu et~al.(2019)Shimizu, Mizusaki, Utsuno, and
  Tsunoda]{SHIMIZU2019372}
Shimizu, N.; Mizusaki, T.; Utsuno, Y.; Tsunoda, Y. Thick-restart block Lanczos method for large-scale shell-model calculations.
\newblock {\em Comput. Phys. Commun.} {\bf 2019}, {\em 244},~372--384.
\newblock {\url{https://doi.org/10.1016/j.cpc.2019.06.011}}.

\bibitem[Wang et~al.(2021)Wang, Huang, Kondev, Audi, and Naimi]{Wang2021}
Wang, M.; Huang, W.; Kondev, F.; Audi, G.; Naimi, S. The AME 2020 atomic mass evaluation (II). Tables, graphs and references.
\newblock {\em Chin. Phys. C} {\bf 2021}, {\em 45},~030003.
\newblock {\url{https://doi.org/10.1088/1674-1137/abddaf}}.

\bibitem[Yoshida et~al.(2018)Yoshida, Utsuno, Shimizu, and
  Otsuka]{PhysRevC.97.054321}
Yoshida, S.; Utsuno, Y.; Shimizu, N.; Otsuka, T.
\newblock Systematic shell-model study of $\ensuremath{\beta}$-decay properties
  and Gamow-Teller strength distributions in $A\ensuremath{\approx}40$
  neutron-rich nuclei.
\newblock {\em Phys. Rev. C} {\bf 2018}, {\em 97},~054321.
\newblock {\url{https://doi.org/10.1103/PhysRevC.97.054321}}.

\bibitem[Stoitsov et~al.(2013)Stoitsov, Schunck, Kortelainen, Michel, Nam,
  Olsen, Sarich, and Wild]{STOITSOV20131592}
Stoitsov, M.; Schunck, N.; Kortelainen, M.; Michel, N.; Nam, H.; Olsen, E.;
  Sarich, J.; Wild, S.
\newblock Axially deformed solution of the
  {S}kyrme-{H}artree-{F}ock-{B}ogoliubov equations using the transformed
  harmonic oscillator basis (II) {HFBTHO} v2.00d: {A} new version of the
  program.
\newblock {\em Comput. Phys. Commun.} {\bf 2013}, {\em 184},~1592--1604.
\newblock {\url{https://doi.org/10.1016/j.cpc.2013.01.013}}.

\bibitem[Abe(2008)]{HFBSPH}
Abe, D.
\newblock{New type of density-dependent effective interaction and its applications to exotic nuclei.}
\newblock{Ph.D. thesis},
\newblock{University of Tokyo},
\newblock{2008}.

\bibitem[Audi and Wapstra(1995)]{AUDI1995409}
Audi, G.; Wapstra, A. The 1995 update to the atomic mass evaluation.
\newblock {\em Nucl. Phys. A} {\bf 1995}, {\em 595},~409--480.
\newblock {\url{https://doi.org/10.1016/0375-9474(95)00445-9}}.

\bibitem[Tilley et~al.(1998)Tilley, Cheves, Kelley, Raman, and
  Weller]{TILLEY1998249}
Tilley, D.; Cheves, C.; Kelley, J.; Raman, S.; Weller, H. Energy levels of light nuclei, \emph{A} = 20.
\newblock {\em Nucl. Phys. A} {\bf 1998}, {\em 636},~249--364.
\newblock {\url{https://doi.org/10.1016/S0375-9474(98)00129-8}}.

\bibitem[Basunia(2015)]{BASUNIA201569}
Basunia, M.S. Nuclear Data Sheets for A = 22.
\newblock {\em Nucl. Data Sheets} {\bf 2015}, {\em 127},~69--190.
\newblock {\url{https://doi.org/10.1016/j.nds.2015.07.002}}.

\bibitem[Basunia and Chakraborty(2022)]{BASUNIA20223}
Basunia, M.S.; Chakraborty, A. Nuclear Data Sheets for A=24.
\newblock {\em Nucl. Data Sheets} {\bf 2022}, {\em 186},~3--262.
\newblock {\url{https://doi.org/10.1016/j.nds.2022.11.002}}.

\bibitem[Basunia and Hurst(2016)]{BASUNIA20161}
Basunia, M.; Hurst, A. Nuclear Data Sheets for A = 26.
\newblock {\em Nucl. Data Sheets} {\bf 2016}, {\em 134},~1--148.
\newblock {\url{https://doi.org/10.1016/j.nds.2016.04.001}}.

\bibitem[{Shamsuzzoha Basunia}(2013)]{SHAMSUZZOHABASUNIA20131189}
{Shamsuzzoha Basunia}, M. Nuclear Data Sheets for A = 28.
\newblock {\em Nucl. Data Sheets} {\bf 2013}, {\em 114},~1189--1291.
\newblock {\url{https://doi.org/10.1016/j.nds.2013.10.001}}.

\bibitem[Basunia and Chakraborty(2024)]{BASUNIA20241}
Basunia, M.S.; Chakraborty, A. Nuclear Data Sheets for A = 30.
\newblock {\em Nucl. Data Sheets} {\bf 2024}, {\em 197},~1--258.
\newblock {\url{https://doi.org/10.1016/j.nds.2024.08.001}}.

\bibitem[Chen and Singh(2016)]{CHEN20161}
Chen, J.; Singh, B. Nuclear Data Sheets for A = 42.
\newblock {\em Nucl. Data Sheets} {\bf 2016}, {\em 135},~1--192.
\newblock {\url{https://doi.org/10.1016/j.nds.2016.06.001}}.

\bibitem[Chen and Singh(2023)]{CHEN20231}
Chen, J.; Singh, B. Nuclear Structure and Decay Data for A = 44 Isobars.
\newblock {\em Nucl. Data Sheets} {\bf 2023}, {\em 190},~1--318.
\newblock {\url{https://doi.org/10.1016/j.nds.2023.06.001}}.

\bibitem[Wu(2000)]{WU20001}
Wu, S.C. Nuclear Data Sheets for A = 46.
\newblock {\em Nucl. Data Sheets} {\bf 2000}, {\em 91},~1--116.
\newblock {\url{https://doi.org/10.1006/ndsh.2000.0014}}.

\bibitem[Chen(2022)]{CHEN20221}
Chen, J. Nuclear Data Sheets for A = 48.
\newblock {\em Nucl. Data Sheets} {\bf 2022}, {\em 179},~1--382.
\newblock {\url{https://doi.org/10.1016/j.nds.2021.12.001}}.

\bibitem[Chen and Singh(2019)]{CHEN20191}
Chen, J.; Singh, B. Nuclear Data Sheets for A = 50.
\newblock {\em Nucl. Data Sheets} {\bf 2019}, {\em 157},~1--259.
\newblock {\url{https://doi.org/10.1016/j.nds.2019.04.001}}.

\bibitem[Pritychenko et~al.(2016)Pritychenko, Birch, Singh, and
  Horoi]{PRITYCHENKO20161}
Pritychenko, B.; Birch, M.; Singh, B.; Horoi, M.
\newblock Tables of E2 transition probabilities from the first 2+ states in
  even–even nuclei.
\newblock {\em At. Data Nucl. Data Tables} {\bf 2016}, {\em
  107},~1--139.
\newblock {\url{https://doi.org/https://doi.org/10.1016/j.adt.2015.10.001}}.

\bibitem[Shimizu et~al.(2017)Shimizu, Abe, Honma, Otsuka, Togashi, Tsunoda,
  Utsuno, and Yoshida]{mcsm_review}
Shimizu, N.; Abe, T.; Honma, M.; Otsuka, T.; Togashi, T.; Tsunoda, Y.; Utsuno,
  Y.; Yoshida, T.
\newblock Monte Carlo shell model studies with massively parallel
  supercomputers.
\newblock {\em Phys. Scr.} {\bf 2017}, {\em 92},~063001.
\newblock {\url{https://doi.org/10.1088/1402-4896/aa65e4}}.

\bibitem[Shimizu et~al.(2021)Shimizu, Tsunoda, Utsuno, and
  Otsuka]{PhysRevC.103.014312}
Shimizu, N.; Tsunoda, Y.; Utsuno, Y.; Otsuka, T.
\newblock Variational approach with the superposition of the symmetry-restored
  quasiparticle vacua for nuclear shell-model calculations.
\newblock {\em Phys. Rev. C} {\bf 2021}, {\em 103},~014312.
\newblock {\url{https://doi.org/10.1103/PhysRevC.103.014312}}.

\bibitem[Tsunoda et~al.(2023)Tsunoda, Shimizu, and
  Otsuka]{PhysRevC.108.L021302}
Tsunoda, Y.; Shimizu, N.; Otsuka, T.
\newblock Shape transition of Nd and Sm isotopes and the neutrinoless
  double-$\ensuremath{\beta}$-decay nuclear matrix element of
  $^{150}\mathrm{Nd}$.
\newblock {\em Phys. Rev. C} {\bf 2023}, {\em 108},~L021302.
\newblock {\url{https://doi.org/10.1103/PhysRevC.108.L021302}}.

\end{thebibliography}


\end{document}